\documentclass[traditabstract]{aa}

\usepackage{natbib}
\usepackage{amssymb,verbatim}
\usepackage{graphicx}
\usepackage{hyperref}
\usepackage{txfonts}

\begin{document}

\title{Non-thermal pressure support in X-COP galaxy clusters}

\author{D. Eckert\inst{1} \and V. Ghirardini\inst{2,3} \and S. Ettori\inst{3,4} \and E. Rasia\inst{5} \and V. Biffi\inst{6,5} \and E. Pointecouteau\inst{7} \and M. Rossetti\inst{8} \and S. Molendi\inst{8} \and F. Vazza\inst{2,9} \and F. Gastaldello\inst{8} \and M. Gaspari\inst{10}\thanks{{\it Einstein} and {\it Spitzer} Fellow} \and S. De Grandi\inst{11} \and S. Ghizzardi\inst{8} \and H. Bourdin\inst{12} \and C. Tchernin\inst{13} \and M. Roncarelli\inst{2}}
\institute{
Max-Planck-Institut f\"{u}r extraterrestrische Physik, Giessenbachstrasse 1, 85748 Garching, Germany\\
\email{deckert@mpe.mpg.de}
\and
Dipartimento di Fisica e Astronomia Universit\`a di Bologna, Via Piero Gobetti, 93/2, 40129 Bologna, Italy
\and
INAF, Osservatorio di Astrofisica e Scienza dello Spazio, via Pietro Gobetti 93/3, 40129 Bologna, Italy
\and
INFN, Sezione di Bologna, viale Berti Pichat 6/2, I-40127 Bologna, Italy
\and 
INAF, Osservatorio Astronomico di Trieste, via Tiepolo 11, I-34131, Trieste, Italy
\and
Dipartimento di Fisica dell’ Università di Trieste, Sezione di Astronomia, via Tiepolo 11, I-34131 Trieste, Italy
\and
 IRAP, Université de Toulouse, CNRS, CNES, UPS, Toulouse, France
\and INAF - IASF Milano, Via E. Bassini 15, 20133 Milano, Italy
\and Hamburger Sternwarte, Universit\"{a}t Hamburg, Gojenbergsweg 112, 40131 Hamburg, Germany
\and Department of Astrophysical Sciences, Princeton University, 4 Ivy Lane, Princeton, NJ 08544-1001, USA
\and INAF-Osservatorio Astronomico di Brera, via E. Bianchi 46, 23807, Merate, Italy
\and Dipartimento di Fisica, Università degli Studi di Roma 'Tor Vergata', via della Ricerca Scientifica 1, 00133 Roma, Italy
\and Center for Astronomy, Institute for Theoretical Astrophysics, Heidelberg University, Philosophenweg 12, 69120, Heidelberg, Germany
}

\abstract{Galaxy clusters are the endpoints of structure formation and are continuously growing through the merging and accretion of smaller structures. Numerical simulations predict that a fraction of their energy content is not yet thermalized, mainly in the form of kinetic motions (turbulence, bulk motions). Measuring the level of non-thermal pressure support is necessary to understand the processes leading to the virialization of the gas within the potential well of the main halo and to calibrate the biases in hydrostatic mass estimates. We present high-quality measurements of hydrostatic masses and intracluster gas fraction out to the virial radius for a sample of 13 nearby clusters with available \emph{XMM-Newton} and \emph{Planck} data. We compare our hydrostatic gas fractions with the expected universal gas fraction to constrain the level of non-thermal pressure support. We find that hydrostatic masses require little correction and infer a median non-thermal pressure fraction of $\sim6\%$ and $\sim10\%$ at $R_{500}$ and $R_{200}$, respectively. Our values are lower than the expectations of hydrodynamical simulations, possibly implying a faster thermalization of the gas. If instead we use the mass calibration adopted by the \emph{Planck} team, we find that the gas fraction of massive local systems implies a mass bias $1-b=0.85\pm0.05$ for SZ-derived masses, with some evidence for a mass-dependent bias. Conversely, the high bias required to match \emph{Planck} CMB and cluster count cosmology is excluded by the data at high significance, unless the most massive halos are missing a substantial fraction of their baryons.}

\keywords{X-rays: galaxies: clusters - Galaxies: clusters: general - Galaxies: groups: general - Galaxies: clusters: intracluster medium - cosmology: large-scale structure}
\maketitle

\section{Introduction}

Galaxy clusters form hierarchically through the merging of smaller halos throughout cosmic time. The gravitational energy released during cluster mergers is responsible for heating the intracluster medium (ICM) to the very high temperatures observed in today's clusters. The kinetic energy developed by merging subunits while falling into the main cluster gets progressively dissipated into heat and the plasma is virialized, thereby forming ever more massive systems \citep[e.g.][]{kravtsov12}. Although we expect that the majority of the energy in today's clusters is virialized, the timescale over which the kinetic energy is dissipated is currently unknown. Hydrodynamical simulations predict that non-thermal processes in the form of turbulence, bulk motions, magnetic fields or cosmic rays contribute at the level of $\sim15-30\%$ to the pressure support in present-day galaxy clusters \citep{lau09,vazza09,battaglia13b,nelson14,shi15,biffi16}. In case of substantial non-thermal pressure, the hydrostatic masses estimated under the assumption that the kinetic energy is fully thermalized should be biased low \citep[e.g.][]{rasia06,nagai07,nelson12,shi16,khatri16}. Non-thermal pressure would also affect the measured thermodynamic properties of the ICM by reducing the fraction of energy that is thermalized \citep{kawa10,fusco13}.

Observationally, the integrated non-thermal pressure fraction is difficult to measure. While the fraction of energy contained within magnetic fields and cosmic rays is known to be small \citep[$\lesssim1\%$, e.g.][]{brunetti14,huber13,ackermann14}, the kinetic energy content in the form of bulk motions and turbulence is currently unknown. Indeed, while the exquisite spectral resolution of the \emph{Hitomi} spacecraft allowed us to assess directly the presence of bulk and turbulent motions for the first time \citep{hitomi16,hitomi17} in the Perseus cluster, setting an upper limit of 4\% on the turbulent-to-thermal energy ratio, the measurement was limited to a small fraction of the cluster's volume in the cluster core, where the non-thermal pressure fraction is expected to be small, especially in relaxed systems. 

An alternative way of estimating the integrated contribution of non-thermal pressure is to look for deviations of ICM quantities (gas fraction, entropy) from the predictions of simple gravitational collapse. In particular, the total baryon fraction of massive clusters is one of the most robust quantities derived in cosmological simulations \citep{white93,evrard97,kravtsov05,mantz14}. Since massive local clusters originate from the collapse of very large regions of the early Universe \citep[$\sim30$ comoving Mpc at $z\sim2$,][]{muldrew15}, their composition should be representative of the Universe as a whole, with little scatter. Recent simulations confirm that while the gas and stellar fractions strongly depend on the adopted baryonic physics (cooling, star formation, feedback from supernovae and active galactic nuclei), the total baryonic fraction of massive clusters is nearly independent of the adopted baryonic setup \citep{planelles13,lebrun14,nifty1,nifty2}. Thus, the baryon fraction of the most massive local clusters can be used to test the validity of the hydrostatic equilibrium assumption and estimate the integrated non-thermal pressure fraction \citep{ghirardini17}.

In this paper, we use high-precision hydrostatic masses obtained from the \emph{XMM-Newton} cluster outskirts project \citep[X-COP,][]{xcop} out to $R_{200}$\footnote{Given an overdensity factor $\Delta$, we define $M_{\Delta},R_\Delta$ as the mass and radius for which $M_{\Delta}/(4/3\pi R_\Delta^3)=\Delta\rho_{\rm crit}$, with $\rho_{\rm crit}=3H(z)^2/8\pi G$.} to estimate the level of non-thermal pressure. We present a high-confidence estimate of the universal gas fraction of galaxy clusters and use our assessment of the universal gas fraction to probe the level of systematics in our hydrostatic mass measurements. We apply the same technique to examine potential systematics in the mass calibration adopted by the \emph{Planck} team to derive cosmological parameters from Sunyaev-Zeldovich (SZ) cluster counts, which has resulted in the well-known tension between cosmic microwave background (CMB) and cluster counts \citep{planck15_24}. In companion papers we present our measurements of the thermodynamic properties of X-COP clusters (Ghirardini et al. 2018) and our high-precision hydrostatic mass estimates (Ettori et al. 2018).

The paper is organized as follows. In Sect. \ref{sec:data} we present the dataset and the methods used to derive gas fraction profiles. In Sect. \ref{sec:method} we estimate the universal gas fraction and describe our method to derive the non-thermal pressure fraction. Our results are presented in Sect. \ref{sec:res} and discussed in Sect. \ref{sec:disc}.

Throughout the paper, we assume a $\Lambda$CDM cosmology with $\Omega_m=0.3$, $\Omega_\Lambda=0.7$ and $H_0=70$ km/s/Mpc. Note that since our clusters are local ($z<0.1$) the results have a very mild dependence on the adopted cosmology. 

\section{Data analysis}
\label{sec:data}

\subsection{The X-COP sample}
\label{sec:sample}

X-COP \citep{xcop} is a very large program on \emph{XMM-Newton} (proposal ID 074441, PI: Eckert) designed to advance our understanding of the physics of the ICM throughout the entire cluster volume. It targets 12 local, massive galaxy clusters selected from the \emph{Planck} all-sky SZ survey. The selected clusters are the most significant \emph{Planck} detections \cite[SNR$>12$ in the PSZ1 sample,][]{psz1} in the redshift range $0.04<z<0.1$. The selected systems span a mass range $3\times10^{14}M_\odot<M_{500}<1.2\times10^{15}M_\odot$. The high signal-to-noise of our clusters in the \emph{Planck} survey allows us to perform a joint analysis of the X-ray and SZ properties of X-COP clusters and to extend our reconstruction of the properties of the ICM out to the virial radius. Detection at the virial radius is achieved both in X-rays and in SZ for 11 out of 12 objects. In the remaining case (A3266), the \emph{XMM-Newton} mosaic does not extend far enough to cover all the way out to $R_{200}$.

In Ghirardini et al. (2018) we present in detail our data analysis technique. Here we briefly summarize the main steps of the procedure.

\begin{itemize}
\item The \emph{XMM-Newton} data are processed using the XMMSASv13.5 software and the extended source analysis software (ESAS) package \citep{snowden08}. Count images, exposure maps, and particle background maps are extracted in the [0.7-1.2] keV band to maximize the source-to-background ratio \citep{ettori10}. To model the contribution from soft protons, we follow the procedure described in \citet{ghirardini17} (see Appendix A). We showed using a large set of $\sim500$ blank-sky pointings that this procedure leads to an overall precision of $\sim3\%$ on the subtraction of the background in the [0.7-1.2] keV band.
\item SZ pressure profiles are extracted from the \emph{Planck} $y$-maps extracted with the MILCA algorithm \citep{hurier13} on the multi-frequency information provided by the high-frequency instrument (HFI) from all the available \emph{Planck} data. The procedure for the extraction of the pressure profiles closely follows \citet{planck5}.
\item \emph{XMM-Newton} spectra are extracted in concentric annuli out to $\sim R_{500}$ using the ESAS package. We apply a full-blown modeling technique to determine the source parameters, simultaneously fitting source and background spectra to determine the contribution of each background component jointly (quiescent particle background cosmic X-ray background, Galactic foregrounds, and soft protons). A single-temperature APEC model \citep{apec} is used to describe the plasma emissivity, leaving temperature, normalization and metal abundance as free parameters. Temperature profiles are then constructed by fitting the full spectral model to the observed spectra. 
\item Gas density profiles are reconstructed from the X-ray maps using the azimuthal median method proposed by \citet{eckert15}. Binned surface-brightness maps are constructed using an adaptive Voronoi tessellation algorithm to ensure a minimum of 20 counts per bin. To reconstruct gas emissivity profiles that are unbiased by the presence of accreting clumps, we determine the median of the surface-brightness distributions in concentric annuli \citep{eckert15}. 
\end{itemize}

For more details on the analysis procedure, we refer the reader to Ghirardini et al. (2018).

\subsection{Hydrostatic masses and gas fraction}
\label{sec:mhse}

Our hydrostatic mass measurements and the procedure to obtain them are described in detail in Ettori et al. (2018). We use as our reference mass model the \emph{backward NFW} model \citep{ettori10}, which describes the mass profile using a Navarro-Frenk-White \citep[NFW,][]{nfw96} model. We assume that the ICM is in hydrostatic equilibrium (HSE) within the potential well and that the kinetic energy has been fully converted into thermal energy, in which case the HSE equation reads
\begin{equation}\frac{dP_{\rm gas}}{dr}=-\rho_{\rm gas}\frac{GM_{\rm HSE}(<r)}{r^2}\end{equation}
with $P_{\rm gas},\rho_{\rm gas}$ the pressure and density of the ICM, $G$ the gravitational constant and $M_{\rm HSE}$ the total hydrostatic mass enclosed within a radius $r$. We use the multiscale technique introduced in \citet{eckert16} to deproject the gas density profile. The concentration and scale radius of the NFW profile and the parameters describing the gas density profile are fit jointly to the measured thermodynamic quantities (X-ray emissivity, spectroscopic X-ray temperature, and SZ pressure) and the global likelihood is sampled using the Markov Chain Monte Carlo (MCMC) sampler \texttt{emcee} \citep{foreman-mackey13}. The high statistical quality of the X-COP data results in relative uncertainties of around 5\% on $M_{\rm HSE}$. In Ettori et al. (2018) we compare our results with several other methods to reconstruct the hydrostatic mass (forward fitting, Gaussian processes) and show that all the methods provide consistent results. We also compare our mass estimates with literature values obtained using weak lensing, caustics and integrated SZ signal and find a good agreement between the various methods.

\begin{figure}
\resizebox{\hsize}{!}{\includegraphics{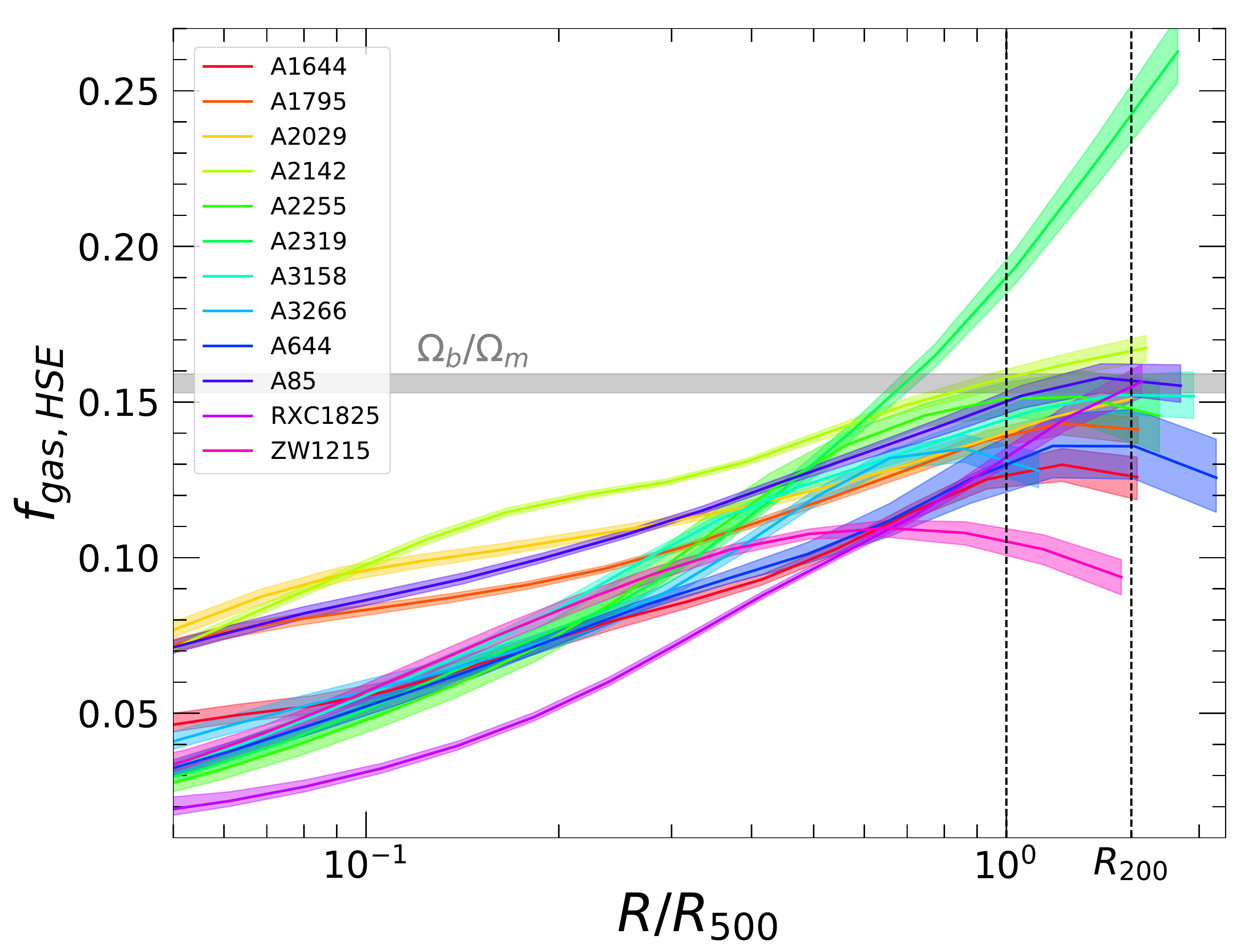}}
\caption{Hydrostatic gas fraction profiles $f_{\rm gas,HSE}(R)=M_{\rm gas}(<R)/M_{\rm HSE}(<R)$ as a function of scale radius $R/R_{500}$ for the X-COP clusters. The gray shaded area shows the \emph{Planck} universal baryon fraction $\Omega_b/\Omega_m$ \citep{planck15_13}.}
\label{fig:fgas_prof}
\end{figure}

At each radius, we integrate the model gas density profiles to determine the enclosed gas mass,
\begin{equation}M_{\rm gas}(<r)=\int_0^r 4\pi r^2 \rho_{\rm gas}(r)\,dr\end{equation}
where $\rho_{\rm gas}=\mu m_{p}(n_e+n_p)$, with $n_e=1.17n_p$ the number densities of electron and proton in a fully ionized gas, $\mu=0.61$ the mean molecular weight, and $m_{p}$ the mass of the proton. The hydrostatic gas fraction profiles are then computed as $f_{\rm gas,HSE}(r)=M_{\rm gas}(<r)/M_{\rm HSE}(<r)$. In Fig. \ref{fig:fgas_prof} we show the hydrostatic gas fraction profiles as a function of the scale radius $R_{500,HSE}$ for the 13 X-COP clusters. The radial range of each profile corresponds to the regions for which information on both the density and the pressure are available. For 10 objects out of 13 our gas fraction profiles extend out to $R_{200}$ without requiring any extrapolation. The typical statistical uncertainties in $f_{\rm gas,HSE}$ are $\sim5\%$ at $R_{500}$ and $\sim10\%$ at $R_{200}$. In the case for which our measurements do not extend all the way out to $R_{200}$ (A3266), the NFW mass model is extrapolated out to $R_{200}$ to estimate the values of $f_{\rm gas,HSE}$.

\section{Methodology}
\label{sec:method}

\subsection{The universal gas fraction}
\label{sec:univ_fgas}

As described in the introduction, the gas fraction of galaxy clusters is a fundamental prediction of cosmological simulations. In this section we review the current knowledge on the universal gas fraction and provide an assessment of the expected true gas fraction. We also conservatively discuss the associated uncertainties.

In a general way, the universal gas fraction within a given radius can be written as 
\begin{equation}
f_{\rm gas,univ}(r)=Y_b(r)\frac{\Omega_b}{\Omega_m}-f_\star,
\label{eq:fgas_univ}
\end{equation}
with $Y_b(r)$  the baryon depletion factor, $\Omega_b/\Omega_m$ the universal baryon fraction, and $f_\star$ the fraction of baryons converted into stars. Baryons in other forms such as warm and molecular gas or dust typically represent less than 0.1\% of the mass content of a galaxy cluster \cite[e.g.][]{edge02,salome03} and for the present study we neglect their contribution. The universal baryon fraction is determined with very high precision by the CMB power spectrum. Here we assume the \emph{Planck} value $\Omega_b/\Omega_m=0.156\pm0.003$ \citep{planck15_13}. This value is slightly lower than, albeit consistent with, the WMAP9 measurement \citep[$0.166\pm0.009$,][]{wmap9}. The baryon depletion factor encodes the fraction of the baryons that is enclosed within a given radius; $Y_b=1$ implies no depletion. We expect from numerical simulation that its value at large cluster-centric radii depends very little on the adopted physical setup, such that the value of $Y_b$ can be calibrated using numerical simulations. Finally, the stellar fraction has been the subject of many studies and its value is well measured. In the following we provide estimates of $Y_b$ and $f_\star$.

\subsubsection{Baryon depletion factor}

The baryon depletion factor $Y_b$ integrated over large fractions of the cluster volume is one of the quantities most robustly predicted by numerical simulations. A recent comparison of 13 different codes including modern and legacy SPH and grid codes \citep{nifty1,nifty2} showed that codes implementing vastly different hydrodynamical solvers and baryonic physics make very consistent predictions on the baryon budget integrated within the virial radius, whereas in the inner regions ($R\lesssim0.5R_{500}$) different codes and simulation setups lead to substantial differences in the predicted baryon fraction. 

In the present work, we utilize simulated clusters with masses in the X-COP range extracted from The Three Hundred Project simulations (hereafter the300, Cui et al. in preparation). 
This project comprises zoom-in re-simulations of more than three hundred Lagrangian regions, of $15$--$20\,h^{-1}$Mpc radius, centered on the most massive cluster-size haloes selected from one of the dark-matter-only MultiDark Simulation run carried out with Planck cosmology \citep{planck15_13}. 
All the regions have been re-simulated at higher resolution (with dark-matter particle mass around $2 \times 10^{9} M_{\odot}$) with a modified version of the Smoothed-Particle-Hydrodynamics GADGET-3 code~\cite[][]{springel05}. The re-simulations include the treatment of a large variety of physical processes to describe the baryonic component, such as gas cooling and star formation, chemical enrichment, stellar and Active Galactic Nuclei (AGN) feedback (see \citealt{rasia15} and references therein for a more detailed description of the hydrodynamical code used, and Cui et al, in prep., for details on the resimulation technique).

\begin{figure}
\resizebox{\hsize}{!}{\includegraphics{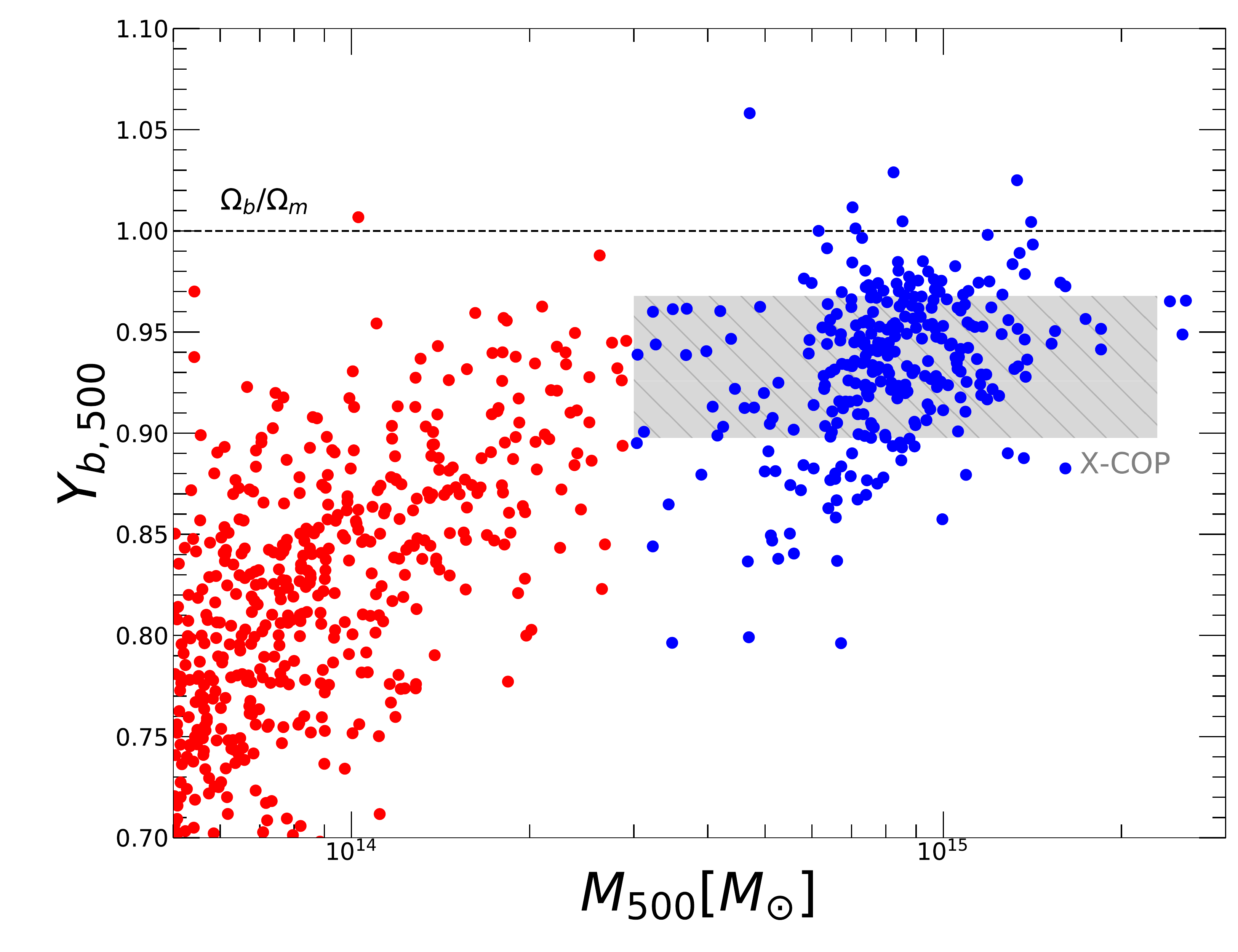}}
\caption{Baryon depletion factor $Y_b$ at $R_{500}$ for the clusters in The300 simulation (filled circles). The blue points show the objects with a mass $M_{500}>3\times10^{14}M_\odot$. The hashed grey shaded area shows the mass range covered by X-COP clusters and our determination of the universal baryon depletion factor and its scatter.}
\label{fig:Yb}
\end{figure}

Thanks to the large statistics afforded by these simulations, we can constrain the value of $Y_b$ and its scatter with good precision. In Fig. \ref{fig:Yb} we show the measurements of $Y_b$ at $R_{500}$ for the whole sample of simulated clusters. As noted in previous studies \citep[e.g.][]{planelles13}, the value of $Y_b$ is typically 5-10\% lower than 1, value that is reached at few times the virial region \citep[see also][]{kravtsov05}. For masses $M_{500} \geq 2\times10^{14}M_{\odot}$, the baryon depletion factor is approximately constant albeit with a large scatter. Smaller systems are instead largely influenced by the active galactic nuclei activity that in one hand pushes outside the hot gas and on the other hand quenches star formation, consequentially reducing both baryonic components.
If we restrict our analysis to the subsample of 295 systems with masses $M_{500}>3\times10^{14}M_\odot$, corresponding to the X-COP mass range, we obtain a median depletion factor of 6.2 per cent at $R_{500}$ and 4.9 per cent at $R_{200}$. The median, $1\sigma$ percentiles, and extreme values of $Y_b$ at $R_{500}$ and $R_{200}$ are provided in Table \ref{tab1}.

\begin{table}
\caption{\label{tab1}Baryon depletion factor in the simulated clusters from the clusters in The300 project with $M_{500}>3\times10^{14}M_\odot$.}
\begin{center}
\begin{tabular}{cccccc}
\hline
$\Delta$ & Median & 16th & 84th & Min & Max\\
\hline
\hline
500 & 0.938 & 0.897 & 0.966 & 0.794 & 1.026\\
200 & 0.951 & 0.923 & 0.982 & 0.875 & 1.024\\
\end{tabular}
\end{center}
\textbf{Column description:} Overdensity $\Delta$; Median value of $Y_b$ in the sample; 16th and 84th percentiles of the values; minimum and maximum values.
\end{table}

\subsubsection{Stellar fraction}
\label{sec:fstar}

While the ICM is known to contain the vast majority of the baryons in galaxy clusters, obviously a fraction of the baryons are locked into stars, both inside galaxies and in the form of intracluster light \citep[ICL,][]{zibetti05,budzynski14,montes14}. Predicting the exact amount of stars in numerical simulations is a difficult endeavour, since the star formation rate and its evolution depend critically on the adopted setup describing gas cooling and feedback from supernovae and AGN \citep[e.g.][]{kravtsov05}. However, the stellar content of galaxy clusters has been extensively studied in the literature \citep{giodini09,gonzalez07,gonzalez13,lagana13,andreon10,chiu17} and the stellar fraction can be robustly set to its observed value. 

\begin{figure}
\resizebox{\hsize}{!}{\includegraphics{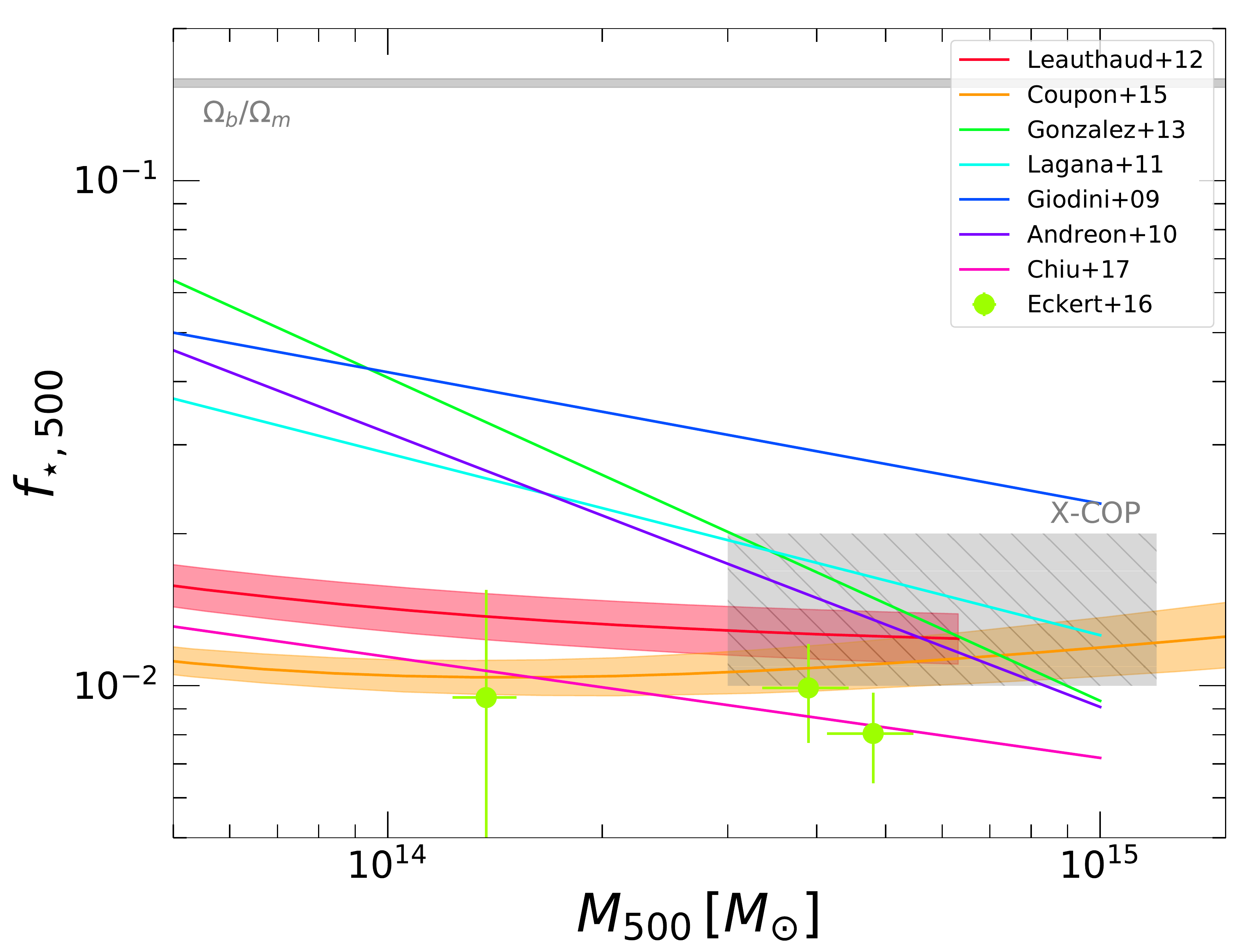}}
\caption{Stellar fraction within $R_{500}$ estimated in several works from the literature. The hashed grey area represents the mass range of X-COP clusters and the range of $f_\star$ adopted in this work. }
\label{fig:fstar}
\end{figure}

In Fig. \ref{fig:fstar} we present a compilation of recent results on the stellar fraction of dark-matter halos as a function of their mass. Results obtained by directly integrating the stellar mass of member galaxies \citep{giodini09,gonzalez13,lagana13,andreon10,chiu17} and from the halo occupation distribution \citep{leauthaud12,coupon15,zu15,eckert16} are compared. While the results obtained with the two methods differ substantially in the galaxy group regime ($M_{500}\lesssim10^{14}M_\odot$), in the mass range covered by X-COP clusters ($3\times10^{14}<M_{500}<1.2\times10^{15}M_\odot$) all studies are broadly consistent and converge to a median stellar fraction of $1.2\%$, with the notable exception of \citet{giodini09}. The contribution of ICL was included in some, but not all cases; measurements indicate that ICL can account for $\sim20-30\%$ \citep{lin04,zibetti05} of the total stellar mass. To encompass the uncertainty associated with the ICL fraction and with the various studies shown in Fig. \ref{fig:fstar}, for the present study we conservatively set the value of the stellar fraction to 
\begin{equation}f_{\star,500}=0.015\pm0.005.\end{equation}
Beyond the central regions the stellar fraction has been shown to be nearly constant \citep{andreon15,vdb15}, thus we adopt the same value for $f_{\star,200}$.

Combining the results from the two previous sections through Eq. \ref{eq:fgas_univ}, we estimate the following values for the universal gas fraction:
\begin{equation}f_{\rm gas,500}=0.131\pm0.009\quad ,\quad f_{\rm gas,200}=0.134\pm0.007.\end{equation}
The errors reported here include both the actual uncertainties in $\Omega_b/\Omega_m$ and $f_\star$ and the measured scatter in $Y_b$.

\subsection{Non-thermal pressure fraction}
\label{sec:ntmet}

Since our X-ray emissivity profiles are corrected for the effects of gas clumping \citep{eckert15} down to the limiting resolution of our observations (10-20 kpc), we expect that residual deviations of the gas mass fraction with respect to the universal gas fraction should be caused by an additional, non-thermal pressure component. In the presence of isotropic non-thermal pressure, the hydrostatic equilibrium equation can be generalized as

\begin{equation} \frac{d}{dr}(P_{th}(r)+P_{NT}(r))=-\rho_{\rm gas}\frac{GM_{\rm tot}(<r)}{r^2},\label{eq:hse}\end{equation}
with $P_{th},P_{NT}$ the thermal and non-thermal pressure components, respectively. We set $\alpha(r)=P_{NT}(r)/P_{\rm tot}(r)$ the non-thermal pressure fraction, i.e. $P_{NT}=\alpha P_{\rm tot}=\frac{\alpha}{1-\alpha}P_{th}$. Using this formulation, Eq. \ref{eq:hse} can be rewritten as
\begin{equation}M_{\rm tot}(<r) = M_{\rm HSE}(<r)+\alpha(r) M_{\rm tot}(<r)-\frac{P_{th}r^2}{(1-\alpha)\rho_{\rm gas} G}\frac{d\alpha}{dr}\end{equation}
with $M_{\rm HSE}(<r)=-\frac{r^2}{\rho_{\rm gas} G}\frac{dP_{th}}{dr}$. The gas fraction as a function of radius can be written as
\begin{eqnarray}f_{\rm gas}(r) & = & \frac{M_{\rm gas}(<r)}{M_{\rm tot}(<r)}\nonumber\\
\, & = & f_{\rm gas,HSE}(r)(1-\alpha)\left(1-\frac{P_{th}r^2}{(1-\alpha)\rho_{\rm gas} GM_{\rm HSE}}\frac{d\alpha}{dr}\right)^{-1}.\label{eq:fgasnt}\end{eqnarray}
Thus, if the true gas fraction is known, the non-thermal pressure fraction $\alpha(r)$ can be estimated by comparing the measured $f_{\rm gas,HSE}$ with the universal value \citep{ghirardini17}. 

For $\alpha(r)$ we use the functional form introduced by \citet{nelson14},
\begin{equation}\frac{P_{NT}}{P_{\rm tot}}(r)=1-A\left(1+\exp\left\{-\left[\frac{r}{Br_{200}}\right]^\gamma\right\} \right)\label{eq:nelson}\end{equation}
with $A,B,$ and $\gamma$ being free parameters. This functional form was shown to reproduce the behavior of the non-thermal pressure fraction in the simulations of \citet{nelson14} and should be approximately valid in the range $[0.3-2]R_{200}$. For the present work, we fix $B=1.7$ to the best fitting value in The300 simulation. Note that in case $\alpha(r)$ is constant this quantity is simply equal to the usual hydrostatic bias $b=1-M_{\rm HSE}/M_{\rm tot}$. 

As already discussed in Sect. \ref{sec:univ_fgas}, the gas fraction predicted by various simulations was found to be consistent \citep{nifty1,nifty2}, including the setup used here \citep[labelled G3X in][]{nifty1}. The predictions however diverge in the inner regions (see their Fig. 10). Thus, we focus on the gas fraction at large radii to compare $f_{\rm gas,HSE}$ to $f_{\rm gas, univ}$ and determine the parameters of $\alpha(r)$. We set the universal gas fraction at $R_{500}$ and $R_{200}$ to the values derived in Sect. \ref{sec:univ_fgas}, and solve numerically Eq. \ref{eq:fgasnt} for the parameters $A$ and $\gamma$. Since this procedure results in a corrected estimate for $M_{500}$ and $M_{200}$, we iteratively repeat the procedure with the revised mass estimates until it converges.

\begin{figure}
\resizebox{\hsize}{!}{\includegraphics{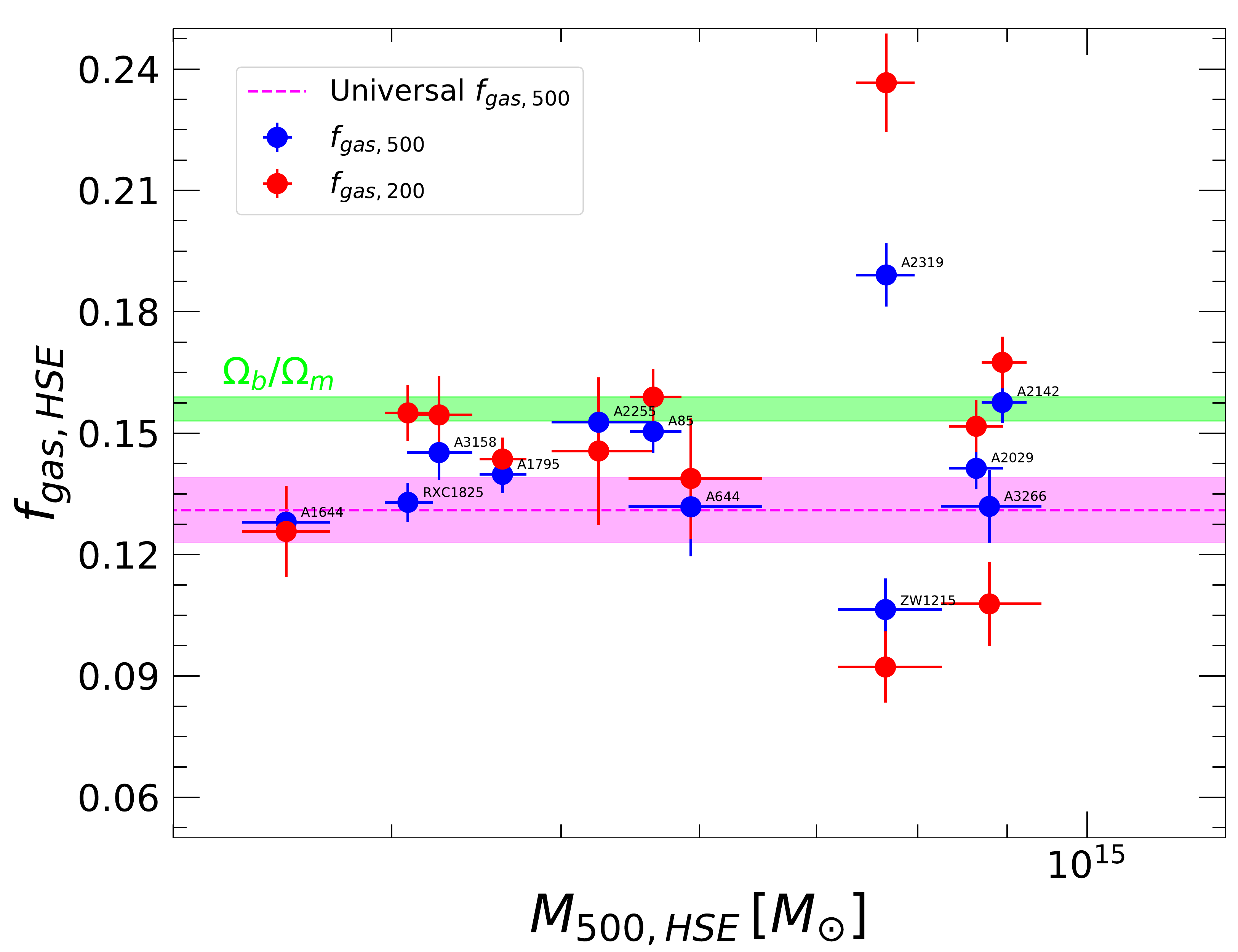}}
\caption{Hydrostatic gas fractions at $R_{500,HSE}$ (blue points) and $R_{200,HSE}$ (red points) obtained from our reference hydrostatic mass model as a function of cluster mass. The dashed magenta line and shaded area represent the universal gas fraction at $R_{500}$ estimated in Sect. \ref{sec:univ_fgas} and its uncertainty. The green shaded area indicates the cosmic baryon fraction \citep{planck15_13}.}
\label{fig:fgashse}
\end{figure}

We use the output MCMC chains of our mass models to propagate the uncertainties in $f_{\rm gas,HSE}$ into our estimate of $\alpha(r)$. We also propagate the dispersion and uncertainties in the universal gas fraction (see Sect. \ref{sec:univ_fgas}) by randomizing the value of $f_{\rm gas}$ in Eq. \ref{eq:fgasnt}. The best-fit curves for $\alpha(r)$ are then computed from the posterior distributions of the parameters.

\section{Results}
\label{sec:res}

\subsection{Non-thermal pressure fraction in X-COP clusters}
\label{sec:nt}

In Fig. \ref{fig:fgashse} we show the values of $f_{\rm gas,HSE}$ for the 13 X-COP clusters at $R_{\rm 500,HSE}$ and $R_{\rm 200,HSE}$ compared to the universal baryon fraction \citep{planck15_13} and the universal gas fraction predicted from The300 simulation. With the exception of A2319 \citep{ghirardini17}, for which a substantial non-thermal pressure support was clearly detected, at $R_{500}$ our measurements of $f_{\rm gas,HSE}$ lie very close to the universal gas fraction, albeit $\sim7\%$ higher on average (median $f_{\rm gas,500}=0.141\pm0.005$, with 12\% scatter), implying a mild contribution of non-thermal pressure. Conversely, at $R_{200}$, the majority of the measurements (9 out of 12) slightly exceed the universal gas fraction (median $f_{\rm gas,200}=0.149_{-0.008}^{+0.006}$). With just two exceptions (A3266 and ZwCl 1215), the gas fraction of all systems is at least as large as our determination of the Universal gas fraction.

To investigate any dependence of the measured gas fraction on the core state, we split our sample into cool core (CC) and non cool core (NCC) classes based on their central entropy $K_0$ as measured by \emph{Chandra} \citep{cavagnolo09} because of its higher resolution in the core, using a threshold of $30$ keV cm$^2$ as a boundary between the two populations. Using this definition, four of our systems are classified as CC, and the remaining eight as NCC. At $R_{500}$, we estimate median values $f_{\rm gas,CC}=0.142\pm0.006$ and $f_{\rm gas,NCC}=0.141\pm0.008$, i.e. there is no difference in the hydrostatic gas fraction of the two sub-populations. The values of $f_{\rm gas, HSE}$ in the NCC population appear to be more scattered than in the CC population (15\% versus $<7\%$). However, the small number of objects in our sample makes it difficult to make any strong statistical claim about the scatter of the two populations. 

We used the distribution of output values for the parameters of the non-thermal pressure fraction (Eq. \ref{eq:nelson}) to determine the non-thermal pressure fraction at $R_{500}$ and $R_{200}$. The normalization of the non-thermal pressure term $A$ in Eq. \ref{eq:nelson} is usually well determined and lies in the range $0.4-0.8$ (median 0.65). The slope $\gamma$ is however poorly constrained, given that we are constraining it using only two anchor points ($R_{500}$, $R_{200}$). In the cases where no modification to the gas fraction was required we computed upper limits at the 90\% confidence level. In Table \ref{tab2} we provide our measurements of the hydrostatic gas fraction, of the non-thermal pressure ratio, and of the total cluster masses after applying the method described in Sect. \ref{sec:ntmet} (labelled as $M_{\rm tot}$ hereafter). The uncertainties on $P_{NT}$ were propagated to the estimated total mass using the MCMC chain (see Sect. \ref{sec:ntmet}). In the cases for which no evidence for non-thermal pressure was found, $M_{\rm tot}$ is just equal to $M_{\rm HSE}$.

\begin{table*}
\caption{\label{tab2}Hydrostatic gas fraction, non-thermal pressure fraction, and total (bias-corrected) masses at $R_{500}$ and $R_{200}$ in X-COP clusters.}
\begin{center}
\begin{tabular}{ccccccccc}
\hline
Cluster & $M_{\rm HSE,500}$ & $M_{\rm HSE,200}$  & $f_{\rm gas,500}$ & $f_{\rm gas,200}$ & $\alpha(R_{500})$ & $\alpha(R_{200})$ & $M_{\rm tot,500}$ & $M_{\rm tot,200}$\\
 & [$10^{14}M_\odot$] & [$10^{14}M_\odot$] & & & [\%] & [\%] & [$10^{14}M_\odot$] & [$10^{14}M_\odot$]\\
\hline
\hline
A1644 & $3.48\pm0.20$ & $6.69\pm0.58$ & $0.128\pm0.008$ & $0.126\pm0.011$ & $<10.5$ & $<14.8$ & $3.52_{-0.22}^{+0.20}$ & $6.58_{-0.59}^{+0.72}$\\
A1795 & $4.63\pm0.14$ & $6.53\pm0.23$ & $0.139\pm0.005$ & $0.144\pm0.005$ & $2.2_{-2.2}^{+5.6}$ & $6.7_{-4.5}^{+6.0}$ & $4.77_{-0.31}^{+0.35}$ & $6.76_{-0.35}^{+0.37}$\\
A2029 & $8.65\pm0.29$ & $12.25\pm0.49$ & $0.141\pm0.005$ & $0.152\pm0.006$ & $6.0_{-5.7}^{+5.8}$ & $10.4_{-10.4}^{+9.0}$ & $8.98_{-0.83}^{+0.84}$ & $13.29_{-0.60}^{+0.78}$\\
A2142 & $8.95\pm0.26$ & $13.64\pm0.50$ & $0.158\pm0.005$ & $0.168\pm0.006$ & $15.8_{-4.8}^{+4.5}$ & $18.6_{-8.8}^{+7.1}$ & $10.50_{-0.89}^{+0.57}$ & $16.37_{-0.82}^{+0.95}$\\
A2255 & $5.26\pm0.34$ & $10.33\pm1.23$ & $0.153\pm0.011$ & $0.146\pm0.018$ & $5.6_{-5.6}^{+6.8}$ & $6.1_{-6.1}^{+6.3}$ & $5.87_{-0.45}^{+0.47}$ & $10.70_{-0.58}^{+0.77}$\\
A2319 & $7.31\pm0.28$ & $10.18\pm0.52$ & $0.189\pm0.008$ & $0.237\pm0.012$ & $43.6_{-3.6}^{+3.5}$ & $52.3_{-4.6}^{+3.4}$ & $11.44_{-1.11}^{+1.06}$ & $20.11_{-1.31}^{+1.14}$\\
A3158 & $4.26\pm0.18$ & $6.63\pm0.39$ & $0.145\pm0.007$ & $0.155\pm0.010$ & $8.5_{-5.8}^{+5.7}$ & $12.5_{-11.6}^{+8.9} $ & $4.53_{-0.37}^{+0.38}$ & $7.34_{-0.35}^{+0.46}$\\
A3266 & $8.80\pm0.57$ & $15.12\pm1.44$ & $0.132\pm0.009$ & $0.108\pm0.018$ & $<11.2$ & $<15.9$ & $8.94_{-0.53}^{+0.60}$ & $14.49_{-2.39}^{+3.01}$\\
A644 & $5.66\pm0.48$ & $7.67\pm0.73$ & $0.132\pm0.012$ & $0.139\pm0.015$ & $3.2_{-3.2}^{+6.4}$ & $5.6_{-5.6}^{+6.4}$ & $6.03_{-0.69}^{+0.62}$ & $8.35_{-0.52}^{+0.70}$\\
A85 & $5.65\pm0.18$ & $8.50\pm0.36$ & $0.150\pm0.005$ & $0.159\pm0.007$ & $10.2_{-5.6}^{+4.9}$ & $11.5_{-9.5}^{+8.9}$ & $6.22_{-0.44}^{+0.54}$ & $9.56_{-0.46}^{+0.53}$\\
RXC1825 & $4.08\pm0.13$ & $6.15\pm0.26$ & $0.133\pm0.005$ & $0.155\pm0.007$ & $5.1_{-5.1}^{+5.1}$ & $15.2_{-7.8}^{+6.4}$ & $3.94_{-0.28}^{+0.36}$ & $6.87_{-0.37}^{+0.40}$\\
ZwCl1215 & $7.66\pm0.52$ & $13.03\pm1.23$ & $0.106\pm0.008$ & $0.092\pm0.009$ & $<11.9$ & $<15.7$ & $7.67_{-0.47}^{+0.59}$ & $13.03_{-1.12}^{+1.37}$\\
\hline
Median & &  & $0.141_{-0.005}^{+0.006}$ & $0.149_{-0.008}^{+0.009}$ & $5.9_{-3.3}^{+2.9}$ & $10.5_{-5.5}^{+4.3}$ & \\
\hline
\end{tabular}
\end{center}
\textbf{Column description:} Cluster name; masses reconstructed using hydrostatic equilibrium (see Ettori et al. 2018); hydrostatic gas fractions; non-thermal pressure ratio $\alpha=P_{\rm NT}/P_{\rm tot}$; total masses corrected for non-thermal pressure. Upper limits are at the 90\% confidence level.
\end{table*}

\begin{figure}
\resizebox{\hsize}{!}{\includegraphics[width=\textwidth]{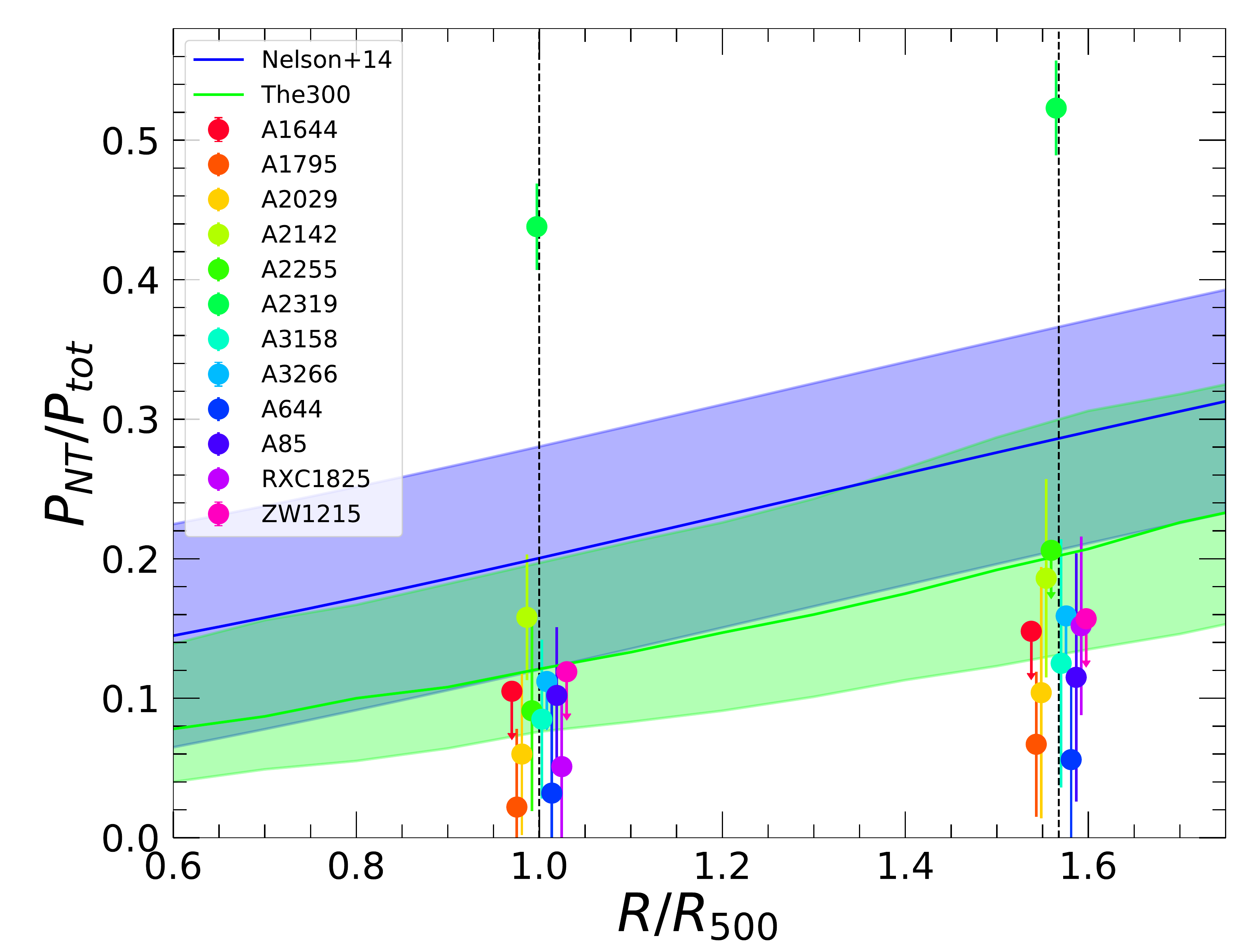}}
\caption{Non-thermal pressure fraction in X-COP clusters at $R_{500}$ and $R_{200}$. The positions on the X axis are slightly shifted for clarity. The blue and green curves and shaded areas show the mean non-thermal pressure ratio predicted from the numerical simulations of \citet{nelson14} and Rasia et al., respectively.}
\label{fig:pnt}
\end{figure}

In Fig. \ref{fig:pnt} we show the non-thermal pressure fractions at $R_{500}$ and $R_{200}$ for the entire X-COP sample. We immediately see that in the vast majority of cases non-thermal support in X-COP clusters is mild. We use a bootstrap method to compute the median of the distribution and the uncertainties on the median. We find a median non-thermal pressure of $5.9_{-3.3}^{+2.9}\%$ and $10.5_{-5.5}^{+4.3}\%$ at $R_{500}$ and $R_{200}$, respectively. For comparison, in Fig. \ref{fig:pnt} we show the average non-thermal pressure ratio in two sets of numerical simulations ($\Omega_{500}$, \citet{nelson14}; The300, Rasia et al. in prep.), with the scatter of the population indicated as the shaded areas. Assuming that the non-thermal pressure support is due to random gas motions, the level of non-thermal pressure in numerical simulations was defined as the ratio of the pressure induced by random motions $P_{\rm rand}=\frac{1}{3}\rho\sigma_{\rm gas}^2$ to the sum of random and thermal pressure \citep{nelson14,biffi16},
\begin{equation}
\frac{P_{NT}}{P_{\rm tot}}=\frac{\sigma_{\rm gas}^2}{\sigma_{\rm gas}^2+(3kT/\mu m_p)},\label{eq:prand}
\end{equation}
with $\sigma_{\rm gas}$ the velocity dispersion of gas particles in spherical shells, $k$ the Boltzmann constant, $\mu$ the mean molecular weight and $m_p$ the proton mass. Interestingly, most of our measurements lie substantially below the \citet{nelson14} curve, possibly indicating a higher level of thermalization in the real population compared to the simulations. A somewhat lower level of non-thermal pressure is predicted in The300 simulation, in better agreement with our results. We discuss this comparison in further detail in Sect. \ref{sec:thermalization}. 

\subsection{Impact of missing clusters and sample selection}

While the selection of the X-COP sample was designed to be fairly clean (see Sect. \ref{sec:sample}), our original selection excluded 4 systems for which we were unsure whether the strategy adopted for the project could be applied. This includes clusters with obvious substructures, aspherical morphology, bad visibility for \emph{XMM-Newton}, or an apparent size barely larger than the \emph{Planck} beam (see Sect. 2.1 of Ghirardini et al. 2018). Two of these systems (A754 and A3667) are extreme mergers which may deviate substantially from hydrostatic equilibrium in a way similar to A2319, thus the average level of non-thermal pressure support in our sample may be biased by the exclusion of these objects. To investigate the potential impact of these systems on our results, we assumed that these two missing objects show a level of non-thermal pressure similar to that of A2319 and that the remaining two are representative of the population. Such a choice has no influence on the median non-thermal pressure fraction, but increases the mean value from 9\% to 13\% at $R_{500}$. We can thus conclude that our analysis sets an upper limit of 13\% to the mean level of non-thermal pressure in the \emph{Planck} cluster population. We were recently allocated observing time on \emph{XMM-Newton} (PI: Ghirardini) to extend the X-COP strategy to the missing objects discussed here, which will allow us to set firm limits on the level of non-thermal pressure support.

Another issue that cannot be addressed with the current X-COP sample is any possible evolution of non-thermal pressure and hydrostatic mass bias with redshift. \citet{nelson14} showed that the expected level of non-thermal pressure is expected to increase substantially with redshift when scaled by the critical density \citep[see also][]{shi16}. By design, the X-COP sample was selected to contain only systems that are well resolved by \emph{Planck}, which limits our study to local ($z<0.1$) systems. Further studies with higher resolution SZ data (e.g. NIKA2, MUSTANG2) will allow us in the near future to test whether the findings reported here can be applied to clusters located at higher redshifts.

\subsection{Comparison with \emph{Planck} SZ masses}
\label{sec:szmass}

Following the discovery of the tension between \emph{Planck} CMB cosmology and SZ number counts \citep{planck15_24}, considerable effort has been devoted to evaluating the accuracy of the mass calibration adopted by the \emph{Planck} collaboration. \emph{Planck} SZ masses were derived from a relation between SZ flux $Y_{SZ}$ and total mass that was calibrated using \emph{XMM-Newton} HSE masses. Biases in the estimation of the mass might arise from the potential impact of non-thermal pressure in the calibration sample, from uncertainties in the calibration of the \emph{XMM-Newton} effective area and/or from the measurement of the total SZ flux from \emph{Planck} data. Cosmological constraints from \emph{Planck} CMB and cluster counts could be reconciled in case the $Y_{SZ}-M_{500}$ relation adopted by the \emph{Planck} team is biased low by a factor $1-b=M_{\rm SZ}/M_{\rm true}=0.58\pm0.04$, presumably because of strong non-thermal pressure support \citep{rasia06,nagai07}. Numerous studies have addressed this issue by directly comparing masses derived using X-ray and weak lensing techniques \citep[e.g.][]{hoekstra15,vdl14,smith15}, resulting in somewhat divergent values for the \emph{Planck} mass bias \citep[$1-b$ in the range 0.7-1.0,][]{sereno15}. Here we take a different route and combine our high-quality measurements of hydrostatic masses with our robust assessment of the universal gas fraction to probe the reliability of \emph{Planck} SZ masses. Indeed, in case SZ masses are incorrect we expect the corresponding gas fractions to deviate from the universal gas fraction, which can be easily tested with our data.

We retrieved the masses of X-COP clusters from the PSZ2 catalog \citep{psz2} and determined the value of $R_{500,SZ}$ accordingly. We recall that the PSZ2 masses were determined by applying a relation between SZ signal $Y_{500}$ and total mass $M_{500,SZ}$ calibrated using HSE masses \citep{planck_11_11}. ZwCl 1215 does not have an associated mass in PSZ2 because of the mask used \citep[point sources and Galactic cuts, see sec 6.3 in][]{psz2}, thus in this case we use the mass provided in PSZ1. Although the SZ mass estimate for ZwCl 1215 is low, there is no contaminating point source at radio or sub-millimeter wavelength that could bias the SZ signal. The overall consistency between the PSZ1 and PSZ2 mass proxy and mass estimates provided by \citet{psz1,psz2} can however not account for features or structures intrinsic to the cluster.

\begin{figure}
\resizebox{\hsize}{!}{\includegraphics{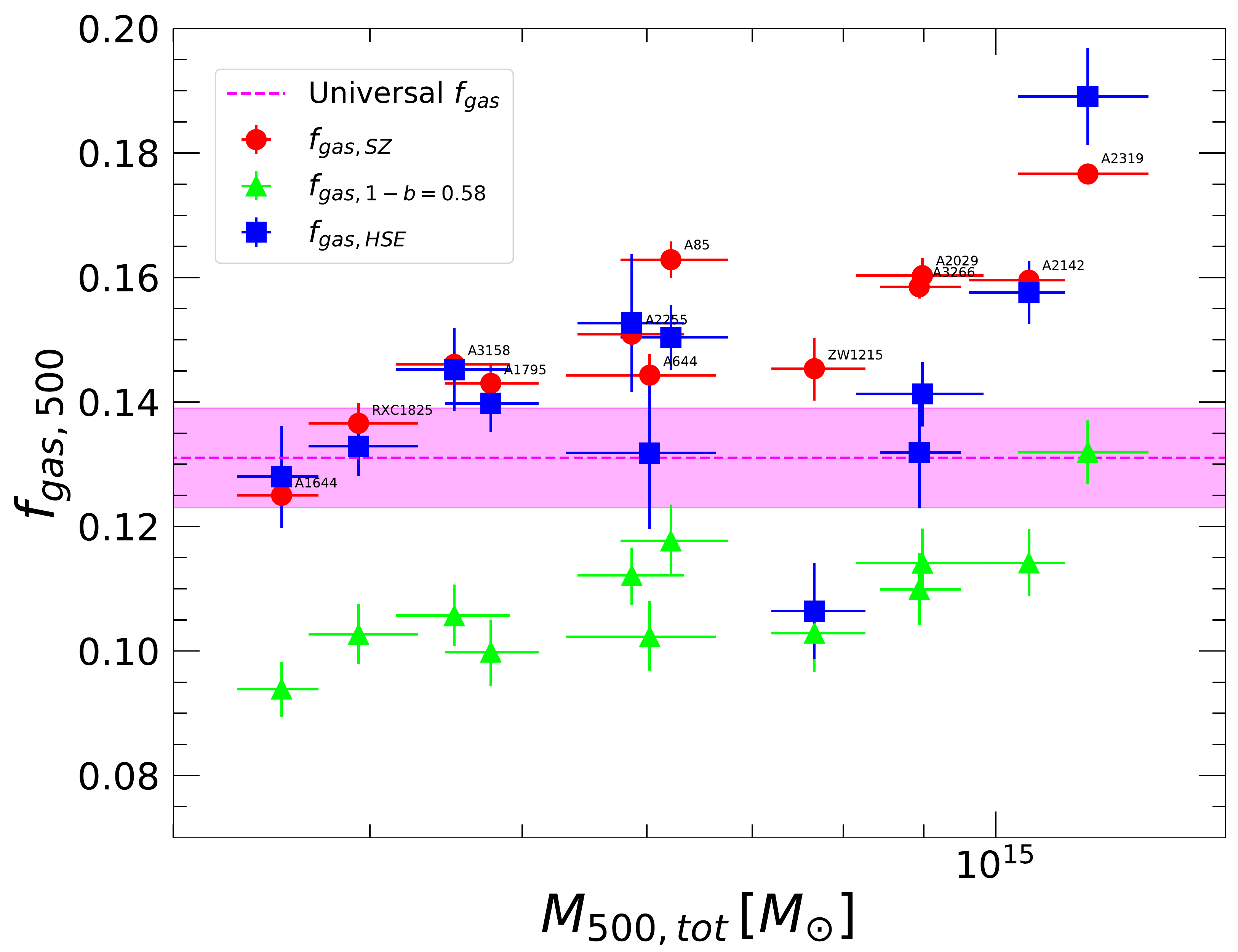}}
\caption{ICM gas fraction at $R_{500}$ obtained from our reference hydrostatic-based mass model (blue squares) as a function of the total mass corrected for non-thermal pressure support (see Sect. \ref{sec:szmass}). The red points show the gas fraction obtained using the \emph{Planck} PSZ2 masses estimated from the $Y_{SZ}-M_{500}$ relation \citep{psz2}, whereas the green triangles show the gas fraction one would get by correcting the PSZ2 masses with a uniform mass bias $1-b=0.58\pm0.04$ \citep{planck15_24}. The dashed magenta line and shaded area represent the universal gas fraction estimated in Sect. \ref{sec:univ_fgas}.}
\label{fig:fgas_sz}
\end{figure}

We integrated our gas masses out to $R_{500,SZ}$ and computed the corresponding values of $f_{\rm gas,SZ}$. We repeated the exercise by correcting the PSZ2 masses assuming a mass bias $1-b=0.58\pm0.04$, and derived the corresponding gas fractions. In Fig. \ref{fig:fgas_sz} we show the gas fractions determined using the hydrostatic equilibrium assumption (see Table \ref{tab2}) as a function of total cluster masses corrected for non-thermal pressure support. We also show the gas fractions $f_{\rm gas,SZ}$ measured from the PSZ2 masses and from the masses corrected to reconcile CMB and SZ number count cosmology. 

In Fig. \ref{fig:fgas_sz} we can clearly see that the gas fraction of X-COP clusters exceeds the expected value in case the \emph{Planck} masses are assumed to be correct. The median gas fraction is $f_{\rm gas,SZ}=0.150_{-0.004}^{+0.006}$, i.e $\sim15\%$ higher than the universal gas fraction.  We also notice a trend of increasing gas fraction with cluster mass, which may indicate a mass-dependent bias. Conversely, when correcting the SZ masses by a factor $1-b=0.58$ the gas fraction is substantially lower than expected, with a median value $f_{\rm gas,1-b=0.58}=0.108\pm0.006$. All objects but one would lie outside of the allowed range for $f_{\rm gas,univ}$. Reconciling CMB and SZ cosmology would thus imply that the most massive local clusters are missing about a third of their baryons.

\begin{figure}
\resizebox{\hsize}{!}{\includegraphics{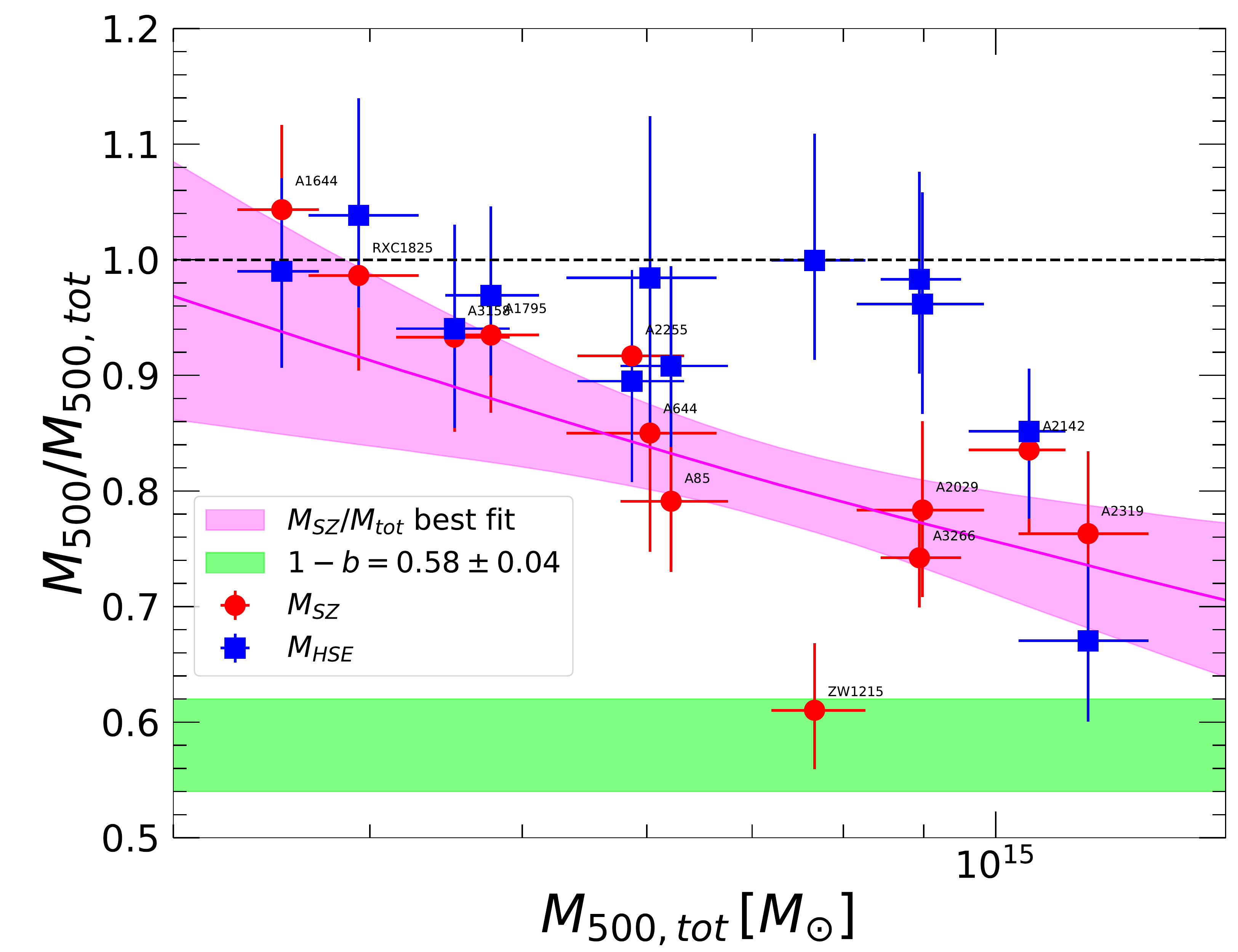}}
\caption{Comparison between HSE (blue squares)/SZ (red dots) and total masses corrected for non-thermal pressure as a function of mass. The magenta line and shaded area show our best fit to the SZ data with a power law, whereas the green area displays the expectation for a constant mass bias $1-b=0.58\pm0.04$.}
\label{fig:bias}
\end{figure}

As shown in Fig. \ref{fig:fgas_sz}, measurements of $f_{\rm gas,500}$ are very sensitive to the adopted mass calibration and thus they can be used to assess systematics in the \emph{Planck} mass calibration. We compared our masses corrected for non-thermal pressure support under the assumption of a universal gas fraction (see Table \ref{tab2}) to the \emph{Planck} SZ-derived masses. In Fig. \ref{fig:bias} we show the ratio of SZ masses to total masses. We measure a median value $1-b=M_{\rm 500,SZ}/M_{\rm 500,tot}=0.85\pm0.05$ for the \emph{Planck} mass bias in our systems. This value may be increased by 3-15\% to take Eddington bias into account \citep{battaglia16}. As noted in several previous studies \citep{vdl14,ettori15}, we observe a substantial mass dependence of the SZ mass bias, with the most massive objects ($M_{500}\sim10^{15}M_\odot$) being biased at the $\sim25\%$ level, while for $M_{500}\sim4\times10^{14}M_\odot$ SZ masses appear to be unbiased. For comparison, in Fig. \ref{fig:bias} we also show the ratio between our direct HSE measurements and the masses corrected for non-thermal pressure support. In the latter case we find that with the notable exception of A2319 our masses require little correction, with a median bias $M_{\rm 500,HSE}/M_{\rm 500,tot}=0.94\pm0.04$.

To assess the dependence of the \emph{Planck} bias on the mass, we describe the relation between SZ mass and total mass as a power law and use the Bayesian mixture model code \texttt{linmix\_err} \citep{kelly07} to fit the data. The resulting parameters read
\begin{equation}\frac{M_{\rm SZ}}{M_{\rm tot}}=\left(0.87\pm0.05\right)\left(\frac{M_{\rm tot}}{5\times10^{14}M_\odot}\right)^{-0.21\pm0.12},\end{equation}
i.e. a mass dependence is detected at $\sim2\sigma$. The best-fitting curve and error envelope are displayed in Fig. \ref{fig:bias}. A high, constant bias $1-b=0.58\pm0.04$ is rejected at the $4.4\sigma$ level for the specific case of the 12 X-COP clusters. Note that the value estimated here for the \emph{Planck} mass bias encompasses all possible kinds of biases in the original \emph{Planck} mass calibration (not restricted to hydrostatic bias), as the gas fractions computed using SZ masses were calculated using the original $M_{500}-Y_{SZ}$ relation.

\section{Discussion}
\label{sec:disc}

\subsection{Systematic uncertainties}

Beyond uncertainties associated with the determination of the universal gas fraction (see Sect. \ref{sec:univ_fgas}), our results can also be affected by potential systematics in our measurements of $f_{\rm gas,HSE}$. Here we review potential sources of systematic uncertainties.

\begin{itemize}
\item \textbf{Reconstruction of $M_{\rm HSE}$: } As described in Sect. \ref{sec:mhse}, we adopt as our reference mass reconstruction method the \emph{backward NFW} method \citep{ettori10}, which assumes that the mass profile can be accurately described by a NFW parametric form. However, this method may be inaccurate if the true mass distribution differs substantially from NFW. In Ettori et al. (2018) we compare the results obtained with our reference \emph{backward NFW} method with the results obtained with methods that do not make any assumption on the shape of the dark matter halo (forward fitting, Gaussian processes). We find that the results obtained with the various methods agree within $\sim5\%$ at a radius of 1.5 Mpc, with the NFW method returning on average slightly higher masses. This propagates to a systematic uncertainty of $\sim5\%$ on the hydrostatic gas fraction, and thus on the non-thermal pressure fraction.
\item \textbf{Gas mass measurements: } The gas mass of local clusters is one of the quantities that can be most robustly computed from X-ray observations. Studies on mock X-ray observations of simulated clusters have shown that measurements of $M_{\rm gas}$ are accurate down to the level of a few percent and exhibit very little scatter, even in situations of violent mergers \citep{nagai07,rasia11,eckert16}. The measured gas densities tend to be biased high in cluster outskirts by the presence of accreting substructures and large-scale asymmetries \citep{mathiesen99,nagai,roncarelli13}, which introduces a systematic uncertainty of 5-10\% on the true gas mass at $R_{200}$. However, thanks to the use of the azimuthal median as a robust estimator of the surface brightness \citep{eckert15}, the bias introduced by infalling substructures has been taken into account in our study. Residual clumping on scales smaller than the resolution of our study ($\lesssim 20$ kpc) can still introduce a slight positive bias in our estimates of $M_{\rm gas}$, however we expect the residual effect to be less than a few percent.
\item \textbf{Calibration uncertainties: } Temperature measurements are known to be affected by systematics of the order of $\sim15\%$ in the high-temperature regime because of uncertainties in the calibration of the effective area of the instrument \citep{nevalainen10,schellenberger15}, with \emph{Chandra} returning systematically higher gas temperatures than \emph{XMM-Newton}. If \emph{Chandra} temperatures are correct, our masses should be underestimated by $\sim15\%$, meaning that our estimates of the non-thermal pressure should be overestimated. However, we note that our mass reconstruction makes use of joint \emph{XMM-Newton} and \emph{Planck} data. In the radial range where data from both instruments are available, we do not observe a systematic offset between X-ray and SZ pressure \citep[see also][]{adam17}. While effective area calibration introduces some uncertainty in the recovered temperature, its effect on the gas density and gas mass is mild. \citet{bartalucci17} compared \emph{XMM-Newton} and \emph{Chandra} reconstructions of gas density profiles and gas masses and found an exquisite agreement between the two missions at the level of 2.5\%. 
\item \textbf{Helium sedimentation: }Several works suggest that Helium nuclei can sediment within cluster cores \citep[e.g.][]{ettori06,markevitch07,peng09}. He abundance can not be directly measured in the ICM, and it is here assumed. We cannot exclude that some sedimentation under the effect of the gravitational potential occurred and is biasing our estimates of the gas mass and total hydrostatic mass \citep[e.g.][]{ettori06}. However, theoretical models predict that any possible rise in the He abundance by up to about 50 per cent over the cosmic value \citep{peng09} induce a bias of a few per cent restricted to the inner ($<R_{2500}$) cluster regions, where sedimentation is more effective, thus not affecting our conclusion at $R\geq R_{500}$.
\end{itemize}

We note that most of the effects discussed here are potential \emph{positive} biases in the reconstruction of $f_{\rm gas,HSE}$. If present, the measured hydrostatic gas fraction is thus more likely to be biased high, which would even lower the required non-thermal pressure and hydrostatic mass bias.

\subsection{Implications on the thermalization of the ICM}
\label{sec:thermalization}

As described in Sect. \ref{sec:nt}, our gas fraction data imply a low level of non-thermal pressure in our population, $\alpha=P_{NT}/P_{\rm tot}=6\%$ and $10\%$ at $R_{500}$ and $R_{200}$, respectively. If we ascribe the excess gas fraction entirely to residual isotropic gas motions (Eq. \ref{eq:prand}), we can relate the measured non-thermal pressure to the velocity dispersion by rewriting Eq. \ref{eq:prand} as
\begin{equation}
\frac{\sigma_{\rm gas}^2}{c_{s}^2}=\mathcal{M}_{\rm 3D}^2(r)=\frac{3}{\gamma}\frac{\alpha(r)}{1-\alpha(r)},
\end{equation}
with $c_{s}=(\gamma kT/\mu m_p)^{1/2}$ the sound speed in the medium, $\mathcal{M}_{\rm 3D}$ the Mach number of residual gas motions, $\gamma=5/3$ the polytropic index, and $\alpha(r)$ the functional form for $P_{NT}/P_{\rm tot}(r)$ following the definition of Eq. \ref{eq:nelson}. The values estimated here thus imply an average Mach number at $R_{500}$
\begin{equation}
\mathcal{M}_{3D,500}=0.33_{-0.12}^{+0.08},
\end{equation}
i.e. isotropic gas motions in the X-COP cluster population are clearly subsonic. This value broadly agrees with the Mach numbers estimated from the amplitude of relative ICM fluctuations \citep{hofmann16,zhu15,eckert17}.

As shown in Fig. \ref{fig:pnt}, our values are somewhat smaller than the predictions of non-radiative adaptive mesh refinement (AMR) simulations by \citet{nelson14} and closer to the curves extracted from the The300 simulation with the smoothed particle hydrodynamics (SPH) code GADGET-3. While in the past legacy SPH codes (employing a typically large artificial viscosity to handle shocks) tended to predict a more clumped and inhomogeneous ICM than grid codes \citep{rasia14}, 
we observe the opposite here. 

A few possible reasons can be given for this difference. First, to better reproduce standard hydrodynamics tests, The300 simulation incorporates a number of advanced features compared to previous SPH codes, including a higher-order Wendland $C^4$ kernel function, the implementation of a time-dependent artificial viscosity scheme, and artificial conduction \citep{beck16}. Compared to previous SPH codes, the SPH scheme included in The300 leads to a more efficient mixing of the gas phases with different entropies. This promotes a faster thermalization of the accreting gas and of small merging substructures, thus reducing the non-thermal pressure fraction. 
On top of that, The300 simulation implements a wide range of baryonic processes (including radiative cooling, star formation, and AGN and supernova feedback) whereas the predictions of \citet{nelson14} are extracted from non-radiative simulations. The balance of cooling and AGN feedback implemented in these simulations substantially changes the appearance of galaxy- and group-scale halos by removing the most structured phase of the ICM from the X-ray emitting phase and by increasing their gas entropy, which leads to smoother and flatter density profiles compared to simulations without powerful feedback mechanisms. The AGN activity provides extra energy to the gas residing in the shallow potential well of small systems, further enhancing its mixing with the cluster ICM during, or immediately after, a merger. 
The subsequent clumping factor is thus reduced compared to the non-radiative case \citep{planelles17} where the entropy difference between the medium and the denser and colder substructure is substantially larger. 
As a result, infalling motions get virialized on shorter timescales and the non-thermal pressure fraction is reduced. 

It should be stressed that the estimate of the non-thermal pressure support in the simulated ICM is by itself non-trivial, owing to the complexity of gas motions in the stratified cluster atmosphere. While all modern simulations overall agree on the predicted radial trend of turbulent motions moving from the cluster centers to the periphery \citep[e.g.][]{vazza11a,nelson14,miniati14,biffi16}, their quantitative answer may change depending on the adopted filtering techniques to disentangle the various velocity components of the ICM (e.g. bulk motions, shock jumps and small-scale chaotic motions),
which is particularly crucial in cluster outskirts \citep[e.g.][]{va17turb,2017MNRAS.470..142S}.
For example, if motions along the radial direction are predominantly directed inwards, the missing pressure estimated with radial averages in simulations may overestimate the non-thermal pressure recovered here using the method devised in Sect. \ref{sec:ntmet}.  More detailed comparisons using exactly the same technique as used here are necessary to test this hypothesis.

\subsection{Implications for cosmology}

The results presented in Sect. \ref{sec:szmass} have important implications for the use of galaxy clusters as cosmological probes. They imply that galaxy cluster masses derived under the assumption of hydrostatic equilibrium in a fully thermalized ICM require little correction from non-thermal pressure support, provided that the 12 X-COP clusters are representative of the \emph{Planck} population. This conclusion is further supported by our direct comparison of hydrostatic and weak lensing masses when available (see Sect. 4 of Ettori et al. 2018), which finds a median ratio $M_{\rm 500,HSE}/M_{\rm 500,WL}=0.87\pm0.10$ and $M_{\rm 200,HSE}/M_{rm 200,WL}=0.86\pm0.13$ for the 6 X-COP clusters with available weak lensing measurements, fully consistent with the non-thermal pressure and the mass ratio $M_{\rm 500,HSE}/M_{\rm 500,tot}=0.94\pm0.04$ estimated from the universal gas fraction method used here.  

At face value, our results strongly disfavor a large hydrostatic bias as the origin of the tension in the $\Omega_m-\sigma_8$ plane between SZ cluster counts and primary CMB. As shown in Fig. \ref{fig:fgas_sz}, our hydrostatic gas fraction measurements are very close to the values obtained with the \emph{Planck} mass calibration, although we note a mildly significant trend of increasing bias in the \emph{Planck} calibration with cluster mass. However, the median mass of the systems in the \emph{Planck} cosmological sample is $\sim5\times10^{14}M_\odot$, where our analysis shows that the SZ masses are biased only at the 10\% level. Although quantifying the exact impact of our results on the cosmological parameters is beyond the scope of this paper, it is fair to say that our study favors lower values of $\sigma_8$ compared to primary CMB, similar to what was obtained from essentially all cluster count \citep{vikhlinin09,dehaan16} and weak lensing tomography studies \citep{heymans13,hildebrandt16}. 

Obviously, the conclusions reached here rest on the premise that our determination of the universal gas fraction is accurate. As shown in Fig. \ref{fig:fgas_sz}, a large, constant hydrostatic bias would imply that the most massive galaxy clusters are missing about a third of their baryons. We also note that our estimate of the stellar fraction (Sect. \ref{sec:fstar}) lies on the high side of the published measurements (see Fig. \ref{fig:fstar}), thus our estimate of $f_{\rm gas}$ is probably on the low side. Extreme AGN feedback would be required to push a substantial fraction of the baryons outside of $R_{200}$, which would lead to high-entropy cores and large offsets from the observed scaling relations \citep[e.g.][]{lebrun14}. High-resolution hydrodynamic simulations testing different AGN feedback models have shown that the feedback must be gentle and tightly self-regulated \citep[e.g.][]{gaspari14}, thus affecting only the regions within $\sim0.1 R_{500}$. An extreme thermal/Sedov blast ($\sim10^{62}$ erg) would be required to evacuate a substantial fraction of the gas away from $R_{500}$, which would transform any CC cluster into a NCC cluster, with cooling times well above the Hubble time. The gentle preservation of many cool cores up to redshift $\sim2$ \citep[e.g.][]{mcdonald17} rules out the strong and impulsive AGN feedback scenario. In the absence of evidence for such extreme phenomena, we conclude for the time being that our estimate of the universal gas fraction does not need to be revised.

\section{Conclusion}
\label{sec:conc}

In this paper, we presented high-precision measurements of the hydrostatic gas fraction from the X-COP project, a sample of 13 clusters with high-quality X-ray and SZ data from \emph{XMM-Newton} and \emph{Planck}. The statistical uncertainties in $f_{\rm gas,HSE}$ are less than 10\% in all cases and measurements at $R_{200}$ are achieved for 10 out of 13 objects without requiring any extrapolation. We used our measurements to estimate the level of non-thermal pressure in our sample. Our results can be summarized as follows.

\begin{itemize}
\item Combining a large set of clusters simulated with a state-of-the-art SPH code with literature measurements of the stellar fraction in observed clusters, we robustly estimate the universal gas fraction of massive clusters to be $f_{\rm gas,500}=0.131\pm0.009$ and $f_{\rm gas,200}=0.134\pm0.007$ at $R_{500}$ and $R_{200}$, respectively. The uncertainties quoted here include both statistical uncertainties and scatter in the simulated cluster population.
\item Our hydrostatic gas fractions are on average consistent with the estimated universal gas fraction, lying just 7\% and 11\% above the universal value at $R_{500}$ and $R_{200}$, respectively, with 12\% scatter. 
\item To determine the integrated level of non-thermal pressure support, we modified the hydrostatic equilibrium equation to incorporate the contribution of a non-thermal pressure term, which we describe using the parametric function of \citet{nelson14} (see Sect. \ref{sec:ntmet}). The parameters of the non-thermal pressure component were then determined by comparing the measured hydrostatic gas fraction profiles with the universal gas fraction. Our procedure leads to revised mass measurements that incorporate the contribution of non-thermal pressure.
\item With the notable exception of A2319 \citep{ghirardini17}, the required levels of non-thermal pressure are mild, with median values $P_{\rm NT}/P_{\rm tot}(R_{500})\sim6\%$ and $P_{\rm NT}/P_{\rm tot}(R_{200})\sim10\%$, with missing clusters possibly raising the non-thermal pressure fraction to a maximum of 13\% at $R_{500}$. These values are lower than the predictions of numerical simulations \citep{nelson14,biffi16}, possibly implying a faster thermalization of the kinetic energy in the real population compared to hydrodynamical simulations.
\item Assuming that the residual non-thermal pressure can be entirely ascribed to random gas motions, we infer an average Mach number $\mathcal{M}_{3D}=0.33_{-0.12}^{+0.08}$, implying that residual kinetic motions are clearly subsonic.
\item We used our masses corrected for the effects of non-thermal pressure to test the accuracy of \emph{Planck} SZ masses in our systems. We find that PSZ2 masses lead to an average gas fraction $f_{\rm gas,SZ}=0.150\pm0.005$ at $R_{500}$, indicating that SZ masses are slightly underestimated. Comparing PSZ2 masses with our masses corrected for non-thermal pressure support, we infer a median bias $1-b=0.85\pm0.05$. As noted in previous studies \citep{ettori15,vdl14}, the bias appears to depend slightly on cluster mass, $M_{\rm SZ}/M_{\rm tot}\propto M_{\rm tot}^{-0.21\pm0.12}$.
\item If instead we assume that the PSZ2 masses are biased low by a constant factor $M_{\rm SZ}/M_{\rm true}=0.58\pm0.04$ to reconcile \emph{Planck} primary CMB and SZ cluster counts, the gas fraction of X-COP clusters would fall short of the universal baryon fraction (median $f_{\rm gas,1-b=0.58}=0.108\pm0.006$), implying that the most massive local clusters would be missing about a third of their baryons. This would pose a serious challenge to our understanding of cluster formation processes and feedback energetics.
\end{itemize}

\begin{acknowledgements}
X-COP data products are available for download on \href{https://www.astro.unige.ch/xcop}{https://www.astro.unige.ch/xcop}. We thank Ricardo Herbonnet and Henk Hoekstra for providing us their weak-lensing results in advance of publication, and the anonymous referee for helpful comments. The research leading to these results has received funding from the European Union’s Horizon 2020 Programme under the AHEAD project (grant agreement n. 654215). Based on observations obtained with XMM-Newton, an ESA science mission with instruments and contributions directly funded by ESA Member States and NASA. S.E. acknowledges financial contribution from the contracts NARO15 ASI-INAF I/037/12/0, ASI 2015-046-R.0 and ASI-INAF n.2017-14-H.0. F.V. acknowledges financial support from the ERC Starting Grant "MAGCOW", no.714196. M.G. is supported by NASA through Einstein Postdoctoral Fellowship Award Number PF5-160137 issued by the Chandra X-ray Observatory Center, which is operated by the SAO for and on behalf of NASA under contract NAS8-03060. Support for this work was also provided by Chandra grant GO7-18121X. E.R acknowledge the ExaNeSt and Euro Exa projects, funded by the European Union’s Horizon 2020 research and innovation programme under grant agreement No 671553 and No 754337 and financial contribution from the agreement ASI-INAF n.2017-14-H.0. H.B. acknowledges financial support by ASI Grant 2016-24-H.0.
\end{acknowledgements}

\bibliographystyle{aa}
\bibliography{NTpressure}

\appendix

\section{The case of A2142}

For one cluster in our sample (A2142), in \citet{tchernin16} we published a similar analysis presenting the gas fraction of this system out to $R_{200}$ based on a joint \emph{XMM-Newton} and \emph{Planck} reconstruction. Noting the convergence of $f_{\rm gas,HSE}$ close to the universal baryon, we concluded that this system requires a modest level of non-thermal pressure, in agreement with the results presented here (15\% and 18\% at $R_{500}$ and $R_{200}$, respectively). Recently, \citet{fusco17} re-analyzed our data and came to the contradicting conclusion that non-thermal pressure support at the level of $\sim30\%$ is required in the outskirts of A2142. To reach this conclusion, \citet{fusco17} neglected the \emph{Planck} constraints while fitting their model and extrapolated their fitted temperature profile beyond the range accessible to \emph{XMM-Newton}. However, as shown in Appendix D of \citet{tchernin16}, the outermost spectral measurement in A2142 is affected by the presence of accreting substructures which bias the measured temperature low. As a result, the extrapolation of the model fitted by \citet{fusco17} underestimates the \emph{Planck} data beyond $R_{500}$ (see their Fig. 1), and their hydrostatic mass reconstruction is biased low. This issue highlights the need to use all the available information in a self-consistent manner.

\end{document}